\documentclass[a4paper,11pt]{article}
\usepackage{pos}

\usepackage{subcaption}
\usepackage{blindtext}
\usepackage{booktabs}
\usepackage{enumitem}
\usepackage{hyperref}
\usepackage{amsfonts}
\usepackage{nicefrac}
\usepackage{caption}
\usepackage{wrapfig}
\usepackage{xcolor}
\usepackage{pifont}
\usepackage{xfrac}
\usepackage{array}
\usepackage{cleveref}
\usepackage[left]{lineno}

\DeclareGraphicsExtensions{.png,.pdf}
\graphicspath{ {./} } 

\title{Measurement of jet yield and acoplanarity using semi-inclusive $\gamma_{\text{dir}}$+jet and $\pi^{0}$+jet distributions in $p$+$p$ and central Au+Au collisions at $\sqrt{s_{\text{NN}}} = 200$ GeV by STAR}

\author*[a]{Derek Anderson (for the STAR Collaboration)}

\affiliation[a]{Cyclotron Institute, Texas A\&M University, \\
  TAMU 3366, College Station, TX 77843}

\emailAdd{derekwigwam9@tamu.edu}

\abstract{The STAR collaboration presents measurements of semi-inclusive distributions of charged jets recoiling from high transverse energy ($E_{\text{T}}$) direct photon and $\pi^{0}$ triggers in $p$+$p$ and central Au+Au collisions at $\sqrt{s_{\text{NN}}} = 200$ GeV.  Jets are reconstructed from charged particles using the anti-$k_{\text{T}}$ algorithm with jet resolution parameters $R = 0.2$ and 0.5.  The large uncorrelated background in central Au+Au collisions is corrected using a mixed-event technique.  This enables a jet measurement extending to low transverse momentum and large $R$ with well-controlled systematic uncertainties.  We present measurements of the jet $R$ dependence of suppression, intra-jet broadening, and acoplanarity of $\pi^{0}$+jet and $\gamma_{\text{dir}}$+jet for trigger $E_{\text{T}}$ between $9 - 20$ GeV.}

\FullConference{%
  41st International Conference on High Energy physics - ICHEP2022\\
  6-13 July, 2022\\
  Bologna, Italy
}

\begin{document}
\maketitle

\section{Introduction}

Heavy-ion collisions at RHIC and the LHC produce a medium of deconfined partons, the Quark-Gluon Plasma (QGP) \cite{QgpRef}.  Hard (large momentum transfer, $Q^{2}$) interactions of quarks and gluons in such collisions generate energetic scattered partons which propagate through the medium and interact with it.  Consequently, the parton showers are modified, a phenomenon known as jet quenching \cite{QuenchRef}.  Jet quenching manifests in several observable effects: transport of energy outside of the reconstructed jet cone, modification of the jet substructure, and enhanced acoplanarity ($\Delta \phi = \phi_{\text{trig}} - \phi_{\text{jet}}$) \cite{AcoplanarityRef}.  While the $\Delta \phi$ distribution has a finite width in vacuum due to Sudakov radiation \cite{SudakovRef}, the presence of a medium may further broaden it due to mechanisms such as multiple in-medium soft scatterings \cite{AcoplanarityRef}, the hard scattering of a parton off QGP quasi-particles \cite{QuasiParticleRef}, and medium response \cite{WakeRef}.

In these proceedings, the STAR collaboration reports measurements of the semi-inclusive yields of jets recoiling from direct photons ($\gamma_{\text{dir}}$) and $\pi^{0}$, together with their acoplanarity distributions in $p$+$p$ and central Au+Au collisions at $\sqrt{s_{\text{NN}}} = 200$ GeV.  Simultaneous measurements of these different observables in the same analysis promise a discriminating and multi-messenger approach to the study of jet quenching.

Since $\gamma_{\text{dir}}$ are color neutral, they do not interact strongly with the QGP; their measured energy thus reflects the $Q^{2}$ of the hard interaction and provides a constraint on the initial energy of the recoiling jet.  Hence, the measurement of jets coincident with a $\gamma_{\text{dir}}$ ($\gamma_{\text{dir}}$+jet) provides a valuable tool for quantifying the effects of jet quenching \cite{WangPhotons}.  In addition, comparison with jets coincident with $\pi^{0}$ ($\pi^{0}$+jet) may elucidate the color factor and path length dependence of medium-induced energy loss, due to differences between the recoil jet populations of the two triggers in their relative quark/gluon fraction and their mean path length \cite{RenkPathLength}.

STAR has previously reported the yield suppression of charged hadrons coincident with $\pi^{0}$ and $\gamma_{\text{dir}}$ triggers \cite{GamHadRef}.  Additionally, STAR has measured the yield of reconstructed charged-particle jets coincident with charged hadron triggers ($h^{\pm}$+jet) using a semi-inclusive approach \cite{HadJetRef}.  In this approach, the large uncorrelated jet background in heavy-ion collisions is corrected with a Mixed Event (ME) technique, enabling the measurement of reconstructed jets at low transverse momentum ($p_{\text{T}}$) and large resolution parameter.  In the current analysis, we combine the $\gamma_{\text{dir}}$/$\pi^{0}$ identification of \cite{GamHadRef} with the semi-inclusive and ME approach of \cite{HadJetRef} to measure the semi-inclusive $\gamma_{\text{dir}}$+jet and $\pi^{0}$+jet yields in $p$+$p$ and central Au+Au collisions.

\section{Analysis}

Two STAR datasets of $\sqrt{s_{\text{NN}}} = 200$ GeV collisions are analyzed: a 10 nb$^{-1}$ sample of Au+Au collisions recorded in 2014, and a 23 pb$^{-1}$ sample of $p$+$p$ collisions recorded in 2009.  Both were recorded using an online high tower trigger, a calorimeter tower above a certain threshold in transverse energy.  Two STAR subsystems are used: the Time Projection Chamber (TPC) \cite{TpcRef}, which provides charged-particle tracks for jet reconstruction, and the Barrel Electromagnetic Calorimeter (BEMC) \cite{BemcRef}, which is used to identify $\pi^{0}$ and $\gamma_{\text{dir}}$ triggers.

Discrimination of $\pi^{0}$ and $\gamma_{\text{dir}}$ candidates in the BEMC is carried out using the Transverse Shower Profile (TSP) method \cite{GamHadRef, TspRef}.  Based on the TSP, the data are separated into two samples: a nearly pure sample of identified $\pi^{0}$, and a sample with an enhanced fraction of $\gamma_{\text{dir}}$ ($\gamma_{\text{rich}}$).

Triggers are selected offline which have a transverse energy ($E_{\text{T}}$) of $E_{\text{T}}^{\text{trig}} = 9 - 20$ GeV and a pseudorapidity of $|\eta^{\text{trig}}| < 0.9$.  The purity of the $\gamma_{\text{rich}}$ sample (the percentage of $\gamma_{\text{rich}}$ that are actually $\gamma_{\text{dir}}$) is determined via a data driven method \cite{GamHadRef, TspRef}.  The $\gamma_{\text{dir}}$+jet distribution is then determined from the $\gamma_{\text{rich}}$ sample via a statistical subtraction, which removes contamination due to hadronic decays and fragmentation photons to the extent that their near-side azimuthal correlations are identical to those of the identified $\pi^{0}$ \cite{GamHadRef, TspRef}.

Jets are reconstructed from the TPC tracks using the anti-$k_{\text{T}}$ algorithm \cite{AntiKtRef, FastJetRef} for two resolution parameters, $R = 0.2$ and 0.5.  Reconstructed jets are subjected to the same fiducial cuts as in \cite{HadJetRef}.  

In Au+Au collisions, there is a substantial background yield of jet candidates which are not correlated with the trigger.  This background yield is removed using the ME technique described in \cite{HadJetRef}.  The uncorrelated jet yield is small in $p$+$p$ collisions, thus no correction for it is applied.  The residual jet transverse momentum ($p_{\text{T}}$) smearing is corrected in two steps \cite{HadJetRef}: first, jets are corrected for an event-wise energy pedestal, and then residual fluctuations caused by detector effects ($p$+$p$ and Au+Au collisions) and the heavy-ion background (Au+Au collisions only) are corrected using regularized unfolding.  We use $p_{\text{T,jet}}^{\text{reco,ch}}$ (where the superscript "ch" denotes "charged jets") to refer to the jet $p_{\text{T}}$ after the event-wise pedestal correction, and $p_{\text{T,jet}}^{\text{ch}}$ to the jet $p_{\text{T}}$ after unfolding.

\begin{figure}
  \centering
  \includegraphics[width=0.9\textwidth]{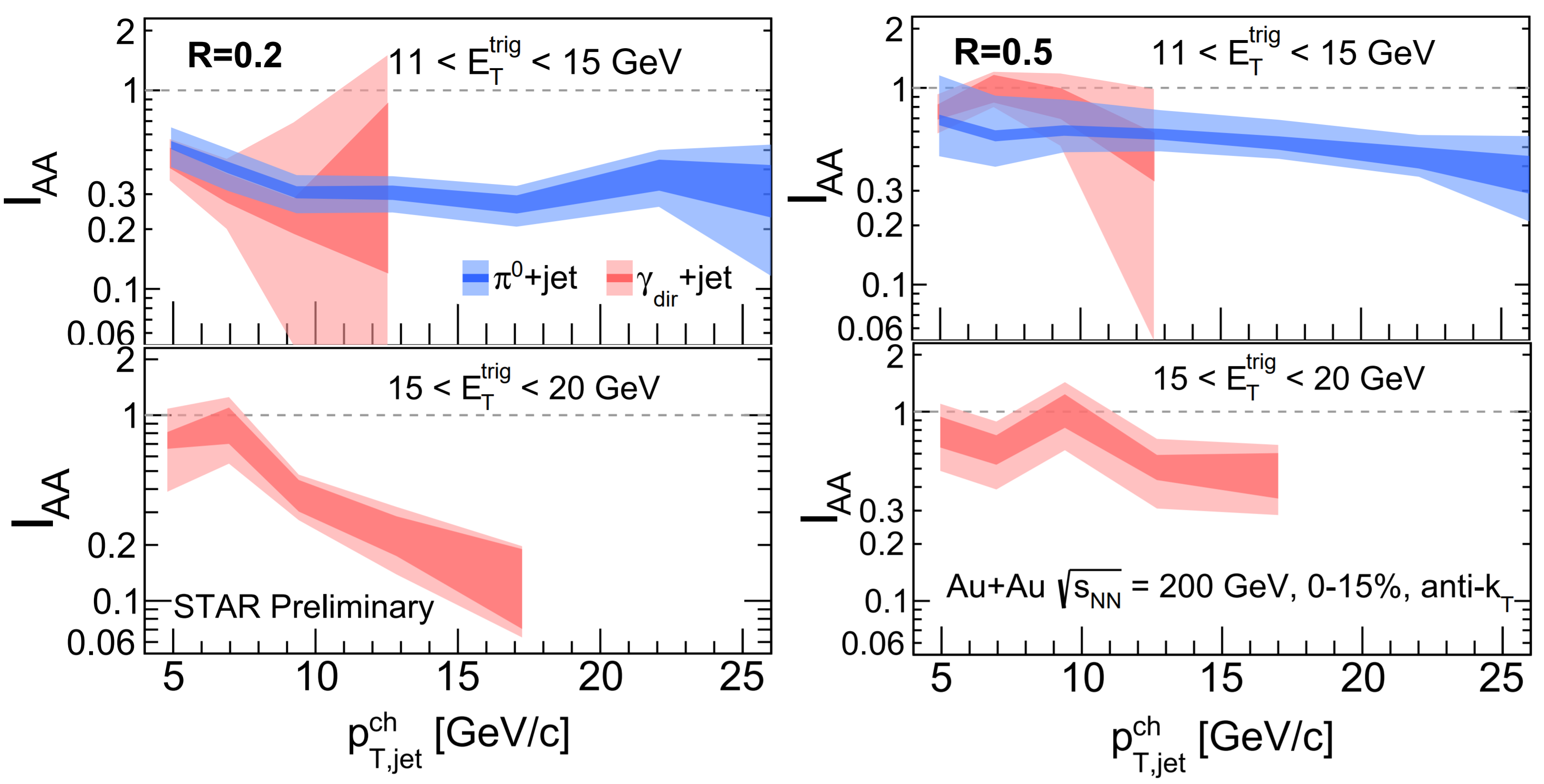}
  \caption{$I_{\text{AA}}$ for $\pi^{0}$+jet (blue) and $\gamma_{\text{dir}}$+jet (red).  Dark bands indicate statistical errors, and light bands indicate systematic uncertainties.}
  \label{Fig:Iaa}
\end{figure}

The two-dimensional acoplanarity distributions must also be unfolded for both $p_{\text{T,jet}}^{\text{reco,ch}}$ and $\Delta \phi$ fluctuations.  Note that the $\Delta \phi$ distributions shown here have been unfolded for $p_{\text{T,jet}}^{\text{reco,ch}}$ fluctuations only.  We estimate that the $\Delta \phi$ smearing effects are small.

\section{Results}

Jet distributions are reported in two ways: the two-dimensional measurement of $\Delta \phi$ vs. $p_{\text{T,jet}}^{\text{ch}}$, and the one-dimensional measurement of $p_{\text{T,jet}}^{\text{ch}}$ for recoil jets, which satisfy $|\Delta \phi - \pi| < \pi / 4$.  The recoil jet $p_{\text{T,jet}}^{\text{ch}}$ distributions in central Au+Au and $p$+$p$ collisions are compared against PYTHIA-8 with the MONASH tune \cite{PythiaRef}.  The PYTHIA-8 distributions are smeared to account for the trigger energy resolution.  We report two different ratios of the trigger-normalized recoil jet yields: $I_{\text{AA}}$, the ratio of the semi-inclusive yield of recoil jets in Au+Au over that in $p$+$p$ for fixed $R$; and $\mathfrak{R}^{0.2/0.5}$, the ratio of the semi-inclusive yield for $R = 0.2$ jets relative to that for $R = 0.5$ jets, for fixed collision system.

Figure \ref{Fig:Iaa} shows $I_{\text{AA}}$ for $E_{\text{T}}^{\text{trig}} = 11 - 15$, $15 - 20$ GeV $\pi^{0}$ and $\gamma_{\text{dir}}$ triggers.  The recoil jet yield for $R = 0.2$ is systematically more suppressed than that for $R = 0.5$.  In addition, the value of $I_{\text{AA}}$ is observed to be consistent within uncertainties between $\pi^{0}$ and $\gamma_{\text{dir}}$ for both values of $R$, despite differences in the recoil jet quark/gluon fraction and mean path length.  Note, however, that the $\gamma_{\text{dir}}$+jet $p_{\text{T,jet}}^{\text{ch}}$ spectrum is steeper, so a similar magnitude of yield suppression corresponds to smaller medium-induced out-of-cone energy loss.

\begin{wrapfigure}{r}{0.5\textwidth}
  \centering
  \includegraphics[width=0.5\textwidth]{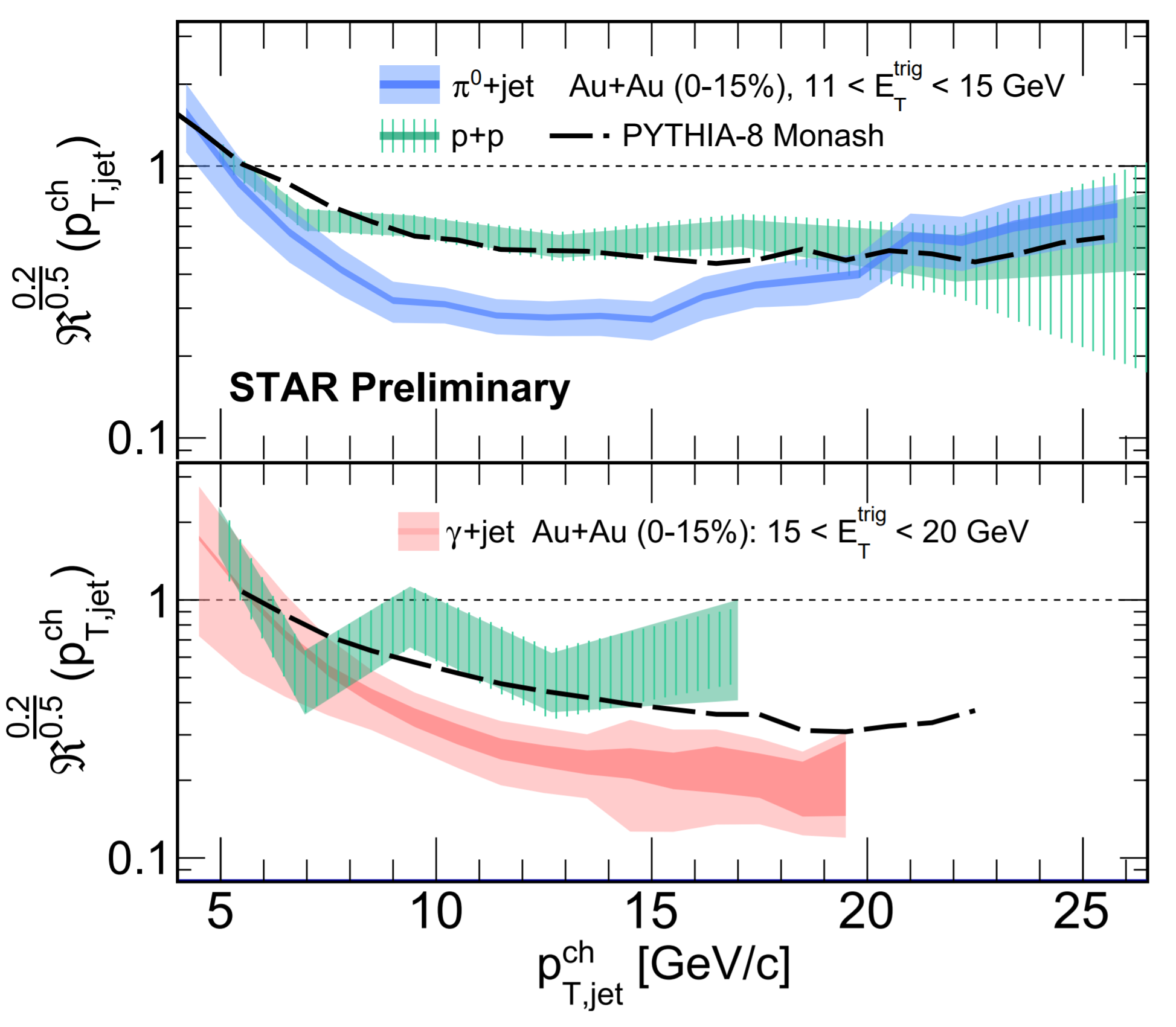}
  \caption{$\mathfrak{R}^{0.2/0.5}$ for $\pi^{0}$ (upper panel) and $\gamma_{\text{dir}}$ (lower panel) triggers from $p$+$p$ (green), Au+Au (blue and red), and PYTHIA-8 (black dashed lines).}
  \label{Fig:R0205}
\end{wrapfigure}

Figure \ref{Fig:R0205} shows the $\mathfrak{R}^{0.2/0.5}$ for $E_{\text{T}}^{\text{trig}} = 11 - 15$ GeV $\pi^{0}$ (upper panel) and $E_{\text{T}}^{\text{trig}} = 15 - 20$ GeV $\gamma_{\text{dir}}$ (lower panel).  We see that $\mathfrak{R}^{0.2/0.5}$ for $p$+$p$ is lower than unity and that PYTHIA-8 reproduces the ratio well.  However, the value of $\mathfrak{R}^{0.2/0.5}$ for central Au+Au is significantly lower than that for $p$+$p$ and PYTHIA-8.

Figures \ref{Fig:Iaa} and \ref{Fig:R0205} show a clear observation of significant medium-induced intra-jet broadening in central Au+Au collisions at RHIC.

\begin{figure}
  \centering
  \begin{subfigure}[t]{0.49\textwidth}
    \includegraphics[width=0.95\textwidth]{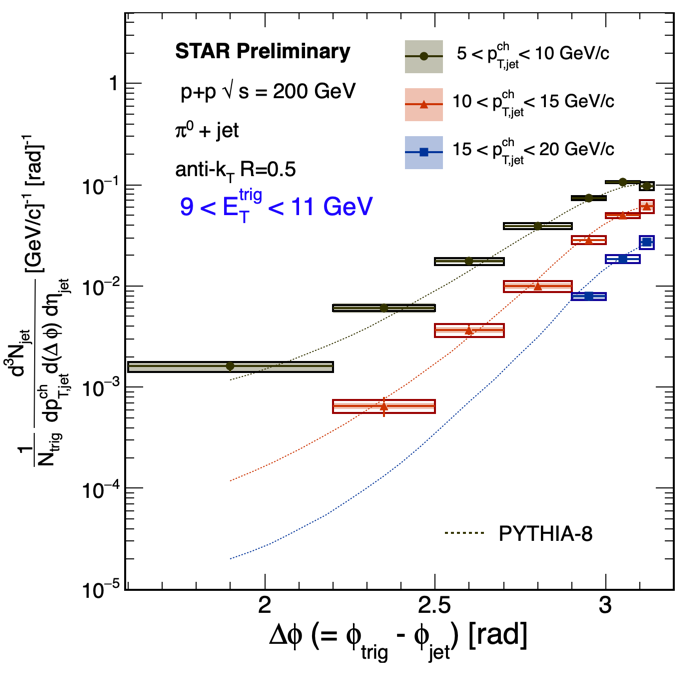}
    \subcaption{}
    \label{Fig:JetDfPP}
  \end{subfigure}
  \begin{subfigure}[t]{0.4825\textwidth}
    \includegraphics[width=0.95\textwidth]{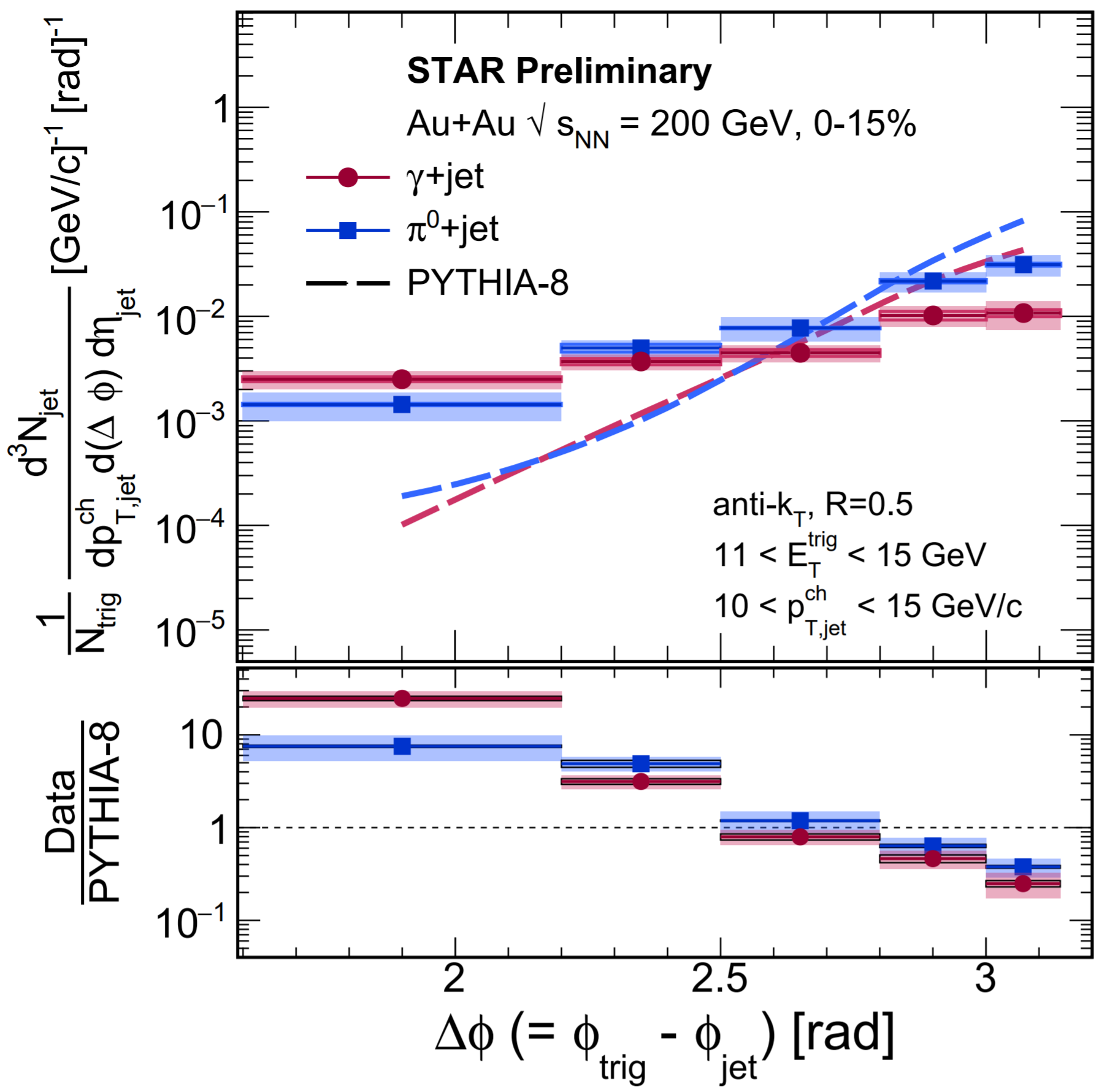}
    \subcaption{}
    \label{Fig:JetDfAA}
  \end{subfigure}
  \caption{Corrected $R = 0.5$ $\Delta \phi$ distributions in $p$+$p$ (a) and Au+Au (b) collisions for $\pi^{0}$ ($p$+$p$ and Au+Au) and $\gamma_{\text{dir}}$ (Au+Au only) triggers.  Vertical lines around data points indicate statistical errors, and filled and open boxes indicate uncorrelated and correlated systematic uncertainties respectively (note that the statistical errors are smaller than the marker size for the Au+Au data points).  Dotted and dashed lines are PYTHIA-8.}
  \label{Fig:JetDf}
\end{figure}

Figure \ref{Fig:JetDfPP} shows the corrected $\Delta \phi$ correlations in $p$+$p$ collisions between $E_{\text{T}}^{\text{trig}} = 9 - 11$ GeV $\pi^{0}$ triggers and $R = 0.5$ jets (boxes).  These distributions are reproduced well by PYTHIA-8 (dotted lines) for all three ranges of $p_{\text{T,jet}}^{\text{ch}}$ ($5 - 10$, $10 - 15$, and $15 - 20$ GeV/$c$).

Figure \ref{Fig:JetDfAA} then shows the corrected $\Delta \phi$ correlations in Au+Au collisions between $E_{\text{T}}^{\text{trig}} =  11 - 15$ GeV $\pi^{0}$ and $\gamma_{\text{dir}}$ triggers and recoil jets of $R = 0.5$ and $p_{\text{T,jet}}^{\text{ch}} = 10 - 15$ GeV/$c$.  The dashed lines are the corresponding distributions from PYTHIA-8, that is validated in the left panel.  We observe a marked enhancement in yield at wide angles (small $\Delta \phi$) in central Au+Au collisions relative to vacuum fragmentation.  This is the first observation of significant medium-induced modification of $\pi^{0}$+jet and $\gamma_{\text{dir}}$+jet acoplanarity at low $p_{\text{T,jet}}^{\text{ch}}$ in central Au+Au collisions at RHIC.

\section{Summary}

STAR has measured the $R$ dependence of recoil jet yield, and acoplanarity using the semi-inclusive distributions of charged-particle jets recoiling from $\pi^{0}$ and $\gamma_{\text{dir}}$ triggers in central Au+Au and $p$+$p$ collisions at $\sqrt{s_{\text{NN}}} = 200$ GeV.  Model calculations based on the PYTHIA-8 event generator are found to be consistent with the measurements in $p$+$p$ collisions.

We have reported both the recoil yield in a fixed angular window as a function of $p_{\text{T,jet}}^{\text{ch}}$, and the distribution of acoplanarity at fixed $p_{\text{T,jet}}^{\text{ch}}$.  We observe marked medium-induced intra-jet broadening.  We also observe clear medium-induced acoplanarity at low jet $p_{\text{T,jet}}^{\text{ch}}$, which may arise from in-medium jet scattering or from the contribution of medium response to the jet signal.  To further investigate the medium-induced acoplanarity and disentangle the underlying mechanisms, it will be essential to extend the kinematic range of this measurement in heavy-ion collisions and compare against theoretical calculations.

\section*{Acknowledgements}

This work is funded in part by the United States Department of Energy under grant number DE-SC0015636.


\end{document}